\documentclass[acmsmall]{acmart}
\usepackage{algorithmic}
\usepackage{algorithm}
\usepackage{multirow}
\usepackage{tabularx}
\usepackage{float}
\usepackage{makecell}
\usepackage{graphicx}

\AtBeginDocument{%
  \providecommand\BibTeX{{%
    \normalfont B\kern-0.5em{\scshape i\kern-0.25em b}\kern-0.8em\TeX}}}

\setcopyright{acmlicensed}
\copyrightyear{2024}
\acmYear{2024}
\acmDOI{XXXXXXX.XXXXXXX}

\acmJournal{JACM}
\acmVolume{37}
\acmNumber{4}
\acmArticle{111}
\acmMonth{8}

\acmISBN{978-1-4503-XXXX-X/18/06}

\begin{document}

\title{EA4RCA: Efficient AIE accelerator design framework for Regular Communication-Avoiding Algorithm}


\author{Wenbo Zhang}
\email{zhangwenbo@bjut.edu.cn}
\orcid{0000-0002-8601-802X}
\affiliation{%
  \institution{Beijing University Of Technology}
  \city{Beijing}
  \country{China}}

\author{Yiqi Liu}
\email{liuyiqi@emails.bjut.edu.cn}
\orcid{0009-0000-8262-2185}
\affiliation{%
  \institution{Beijing University Of Technology}
  \city{Beijing}
  \country{China}}

\author{Tianhao Zang}
\email{zth_bjutnew@emails.bjut.edu.cn}
\orcid{0009-0000-6514-8266}
\affiliation{%
  \institution{Beijing University Of Technology}
  \city{Beijing}
  \country{China}}

\author{Zhenshan Bao}
\email{baozhenshan@bjut.edu.cn}
\orcid{0000-0003-4714-8312}
\affiliation{%
  \institution{Beijing University Of Technology}
  \city{Beijing}
  \country{China}}

\renewcommand{\shortauthors}{W. Zhang et al.}

\begin{abstract}
  With the introduction of the Adaptive Intelligence Engine (AIE), the Versal Adaptive Compute Acceleration Platform (Versal ACAP) has garnered great attention. However, the current focus of Vitis Libraries and limited research has mainly been on how to invoke AIE modules, without delving into a thorough discussion on effectively utilizing AIE in its typical use cases. As a result, the widespread adoption of Versal ACAP has been restricted. The Communication Avoidance (CA) algorithm is considered a typical application within the AIE architecture. Nevertheless, the effective utilization of AIE in CA applications remains an area that requires further exploration. We propose a top-down customized design framework, \textbf{EA4RCA}(Efficient AIE accelerator design framework for regular Communication-Avoiding Algorithm), specifically tailored for CA algorithms with regular communication patterns, and equipped with AIE Graph Code Generator software to accelerate the AIE design process. The primary objective of this framework is to maximize the performance of AIE while incorporating high-speed data streaming services. Experiments show that for the RCA algorithm Filter2D and Matrix Multiple (MM) with lower communication requirements and the RCA algorithm FFT with higher communication requirements, the accelerators implemented by the RA4RCA framework achieve the highest throughput improvements of 22.19x, 1.05x and 3.88x compared with the current highest performance acceleration scheme (SOTA), and the highest energy efficiency improvements of 6.11x, 1.30x and 7.00x.
\end{abstract}


\begin{CCSXML}
<ccs2012>
   <concept>
       <concept_id>10010520.10010521.10010542.10010546</concept_id>
       <concept_desc>Computer systems organization~Heterogeneous (hybrid) systems</concept_desc>
       <concept_significance>500</concept_significance>
       </concept>
 </ccs2012>
\end{CCSXML}

\ccsdesc[500]{Computer systems organization~Heterogeneous (hybrid) systems}

\keywords{AI Engine,  Communication avoiding, Hardware accelerator, Heterogeneous computing, Versal ACAP}

\received{29 March 2024}
\received[revised]{3 May 2024}
\received[accepted]{2 July 2024}
\maketitle

\section{Introduction}

The AMD/Xilinx Versal ACAP architecture \cite{versal1}, which takes AI Engine (AIE) \cite{versal2,versal3} as the core, has become the representative of a new generation of heterogeneous computing architecture. It effectively integrates processing system (PS), programmable logic (PL) and AIE. Compared with the high-performance computing architecture widely used in the current industry, AIE can provide customized computing engine design, and configure flexible and convenient high-speed network of chips (NoC) \cite{versal4,versal5}, so it can implement the customization of accelerators according to the application needs of software and hardware, so as to effectively control the system performance. Although these advances have been made at the architecture level, many of the current solutions only provide a deployment method based on ACAP, and their performance is not outstanding. They do not discuss accelerator design ideas from the perspective of AIE's architecture characteristics, sacrificing the customization characteristics that should belong to ACAP, such as configurability and adaptability, and re customization development is too time-consuming. Therefore, the emergence of AIE has not attracted more accelerator designers' attention. Some work that could highlight AIE's characteristics is still doing relevant work in GPU \cite{gpu1,gpu2,gpu3} or traditional FPGA\cite{fpga1,fpga2,fpga3,fpga4,fpga5,fpga6}.

To leverage the capabilities of AIE, both AMD/Xilinx and numerous researchers have started to develop dedicated applications focused on AIE \cite{vitis_lib, acap_impl1, acap_impl2, acap_impl3, acap_impl4, acap_impl5, acap_impl6, acap_impl7, acap_impl8, acap_impl9}. For instance, A-U3D \cite{acap_impl2} utilizes 96 AIE cores to enable 2D standard convolution, Jie Lei et al. \cite{acap_impl4} have designed an efficient single AIE core for the GEMM algorithm, and Nick Brown \cite{acap_impl7} has employed 60 AIE cores for atmospheric advection simulation, thereby expanding the application landscape of AIE to varying extents. However, these studies have been constrained by application-specific characteristics or the developmental challenges on the PL end, resulting in the utilization of only a small fraction of available AIE cores (<25\%). Some notable exceptions with higher AIE utilization rates include CHARM \cite{acap_impl3}, which employs 384 AIE cores (96\%) for implementing GEMM operators, and XVDPU \cite{acap_impl1}, which utilizes 256 AIE cores (64\%) to achieve efficient convolution. Although these works achieve significant parallelism, they are typically heavily tailored to specific applications, making it difficult to abstract the design principles and apply them to different application types with ease.

In 1984, James Smith proposed the Decoupled Access Execute Architecture (DAE) \cite{DAE}, which addressed the highly decoupled access and execution of operands at a fine-grained hardware level, resulting in significant performance improvements. Furthermore, researchers have integrated DAE into FPGA designs \cite{DAE_FPGA_IMPL}, as FPGA design directly pertains to hardware implementation. Subsequently, Professor James Demmel of Berkeley introduced Communication Avoiding (CA) \cite{CA1}, which decouples at a higher algorithmic level compared to DAE, enabling the widespread deployment of CA algorithms in GPU, NPU, and other architectures for high-performance computing. AIE belongs to the same category as GPU and NPU but exhibits a coarser design granularity in comparison to FPGA and pure hardware accelerator designs. Therefore, it is crucial to discuss an efficient accelerator design framework for CA applications in ACAP architecture. By applying CA principles to regularize and decouple the computing and communication operations of AIE, as well as organizing communication data streams, we can maximize the utilization of hardware resources while accommodating diverse applications.

The ACAP architecture itself offers superior performance and energy efficiency capabilities compared to GPUs. The benefits of AIE, such as flexible configuration and high performance, are theoretically evident due to their applicability in current high-performance applications. However, based on the current development status, only a limited number of works fully leverage these hardware advantages, which hampers the migration of existing high-performance applications to this platform. Consequently, we are motivated to propose the EA4RCA framework based on the following three factors.

\textbf{Motivation  1}: Effectively abstracting application strategies for different hardware based on the characteristics of the ACAP architecture is essential. Given that ACAP comprises heterogeneous hardware, it is necessary to decompose and map upper-level applications to different hardware components to ensure efficient execution. Therefore, it is crucial to propose a well-defined mapping and organizational strategy that optimizes the utilization of the available hardware resources.

\textbf{Motivation  2}: The absence of a flexible configuration scheme poses challenges to efficiently organize the operation of a large number of AIE cores. Currently, many existing AIE deployment schemes struggle to achieve high AIE utilization rates, thereby hindering the realization of optimal performance.

\textbf{Motivation  3}: The full utilization of the hierarchical and fine-grained configuration feature of the NOC data flow remains incomplete, necessitating the development of a more efficient general-purpose data engine tailored to specific application requirements.   While the official PL Data Mover demonstrates satisfactory performance, its functionality is limited.   Currently, numerous tasks are constrained by the development and compilation capabilities of PL. However, the customized data engine encounters issues, such as inadequate performance.

To enhance the utilization of ACAP architecture, this paper introduces the EA4RCA framework tailored for the application of Regular Communication-avoiding Algorithm (RCA). The inherent nature of RCA application allows for straightforward separation of communication and computation. By harnessing the exceptional capabilities of AIE, the EA4RCA framework exhibits remarkable performance advantages in effectively addressing these challenges. The contributions of this research are outlined as follows:

\begin{itemize}
\item  \textbf{The EA4RCA framework introduces a regular communication design pattern that focuses on enhancing computational efficiency.} By abstracting the mapping relationship between hardware and applications, EA4RCA facilitates a "top-down" design process, starting from application characteristics. This approach enables developers to swiftly obtain optimal and viable solutions, thereby effectively enhancing both accelerator performance and development efficiency.
\item  \textbf{AIE configuration design method and automatic code generation based on EA4RCA framework.} This method effectively manages a large number of AIE cores while preserving high flexibility.  Additionally, we introduce the AIE Graph code generator, which simplifies the process of generating the entire AIE design project by simply importing the configuration file.
\item  \textbf{High performance data service method for AIE.}The characteristics of PL highly customized hardware in ACAP architecture were fully utilized to make PL meet the large number of data exchange requirements of AIE.
\end{itemize}

This paper is structured as follows: Chapter 2 provides an introduction to the background and previous research in the field. Chapter 3 presents the comprehensive framework's overall structure and provides detailed explanations of the computing engine, data engine, and AIE Graph code generator. In Chapter 4, we utilize this framework to design accelerators for four applications and evaluate the performance of the EA4RCA framework. Finally, Chapter 5 concludes the paper.

\section{BACKGROUND}
\subsection{Versal ACAP architecture}
With the gradual deepening of researchers' understanding of parallel architecture, it is more and more clear to conclude that efficient parallel architecture not only pursues the operation efficiency of multiple cores, but also needs to flexibly configure a variety of different types of processing cores according to the application characteristics, and also needs to be equipped with corresponding data streams. Based on the above considerations, AMD Xilinx proposed an architecture ACAP with adaptive and highly reconfigurable characteristics\cite{versal1}.

The VCK5000 Versal Development Card is based on the Xilinx 7 nm Versal ACAP architecture and features the first generation 8x50 2D AIE core array. Each AIE core is equipped with a highly efficient Very Long Instruction Word (VLIW) Single Instruction Multiple Data (SIMD) vector processor, capable of executing vector operations up to 1024 bits per cycle at a frequency of 1.33GHz. Furthermore, the programmable logic (PL) component can be designed as dedicated hardware with specific data processing capabilities to serve the AIE. The VCK5000 card is equipped with a 16GB DDR on-board memory, offering a peak bandwidth of 102.4GB/s.

To leverage the performance advantages of ACAP and further advance AIE utilization, AMD/Xilinx officials have provided an extensive collection of library functions in Vitis Libraries\cite{vitis_lib}. Additionally, in 2023, a customized computing challenge (CCC 2023)\cite{CCC2023} was conducted, focusing on two applications: Filter2D, characterized by low communication, and FFT, characterized by high communication. While numerous studies have explored the deployment of related accelerators on various platforms\cite{fft_impl1,fft_impl2,fft_impl3,fft_impl4,fft_impl5,TACO2_GPUFFT}, limited research exists on deploying these specific types of accelerators on AIE, resulting in low resource utilization. For instance, the FFT accelerator proposed by the CCC2023 runner-up team utilizes only 2.25\% of the AIE core, while the Filter2D accelerator proposed by the champion team, which exhibits the highest hardware utilization rate, utilizes 13.5\% of the AIE core. Moreover, the library functions offered by AMD/Xilinx primarily focus on single-core implementations, resulting in a utilization rate of less than 1\%. Thus, the key to addressing this issue lies in proposing a design framework that effectively harnesses the powerful hardware resources of ACAP.

\subsection{Decoupling of computation and communication}

In order to enhance the computational efficiency of algorithms on computers, numerous prior studies have acknowledged the significance of effectively decoupling and partitioning communication and computation as a means to improve performance. It is crucial to minimize the interference caused by communication during the computation process, as it can lead to varying degrees of reduction in computational efficiency.

In 1984, James Smith initially proposed the concept of DAE\cite{DAE}for designing the underlying hardware architecture. This approach achieves a high level of decoupling between operand access and execution, enabling processors to mitigate communication interference during design and exhibit characteristics of communication-computing decoupling. Taking advantage of the notable features of DAEs, researchers have applied them to FPGA\cite{DAE_FPGA_IMPL}and various hardware accelerators\cite{DAE_OTHER_IMPL}. Plasticine\cite{Plasticine} also adopts a design approach similar to communication aggregation to improve performance.

Following James Smith's work, Professor James Demmel introduced the CA algorithm\cite{CA1} for upper-layer algorithms. From the perspective of application programs, this algorithm effectively reduces inter-layer communications and operates at a higher level of abstraction. Moreover, the CA algorithm imposes a certain distribution pattern on the communication and computation of the algorithm, enabling processors to effectively minimize communication overhead and focus on computational operations. The CA algorithm gained popularity\cite{CA2,CA3,CA4,CA5,CA6} in various high-performance computing applications, becoming a key method in parallel system design. RCA applications, which exhibit a regular relationship between computation and communication, offer effective decoupling of communication and computation and can be categorized into two types based on traffic: low communication and high communication. Low communication applications are less susceptible to communication interference during computation, such as MM\cite{TAC01_MM} Filter2D. On the other hand, high communication applications, exemplified by butterfly operations\cite{butterfly2} like the fast Fourier transform (FFT)\cite{butterfly1}, require communication to be as regular as possible during accelerator design.

\section{EA4RCA framework}

\subsection{EA4RCA framework architecture}

\begin{figure}[t]
  \centering
  \includegraphics[width=\linewidth]{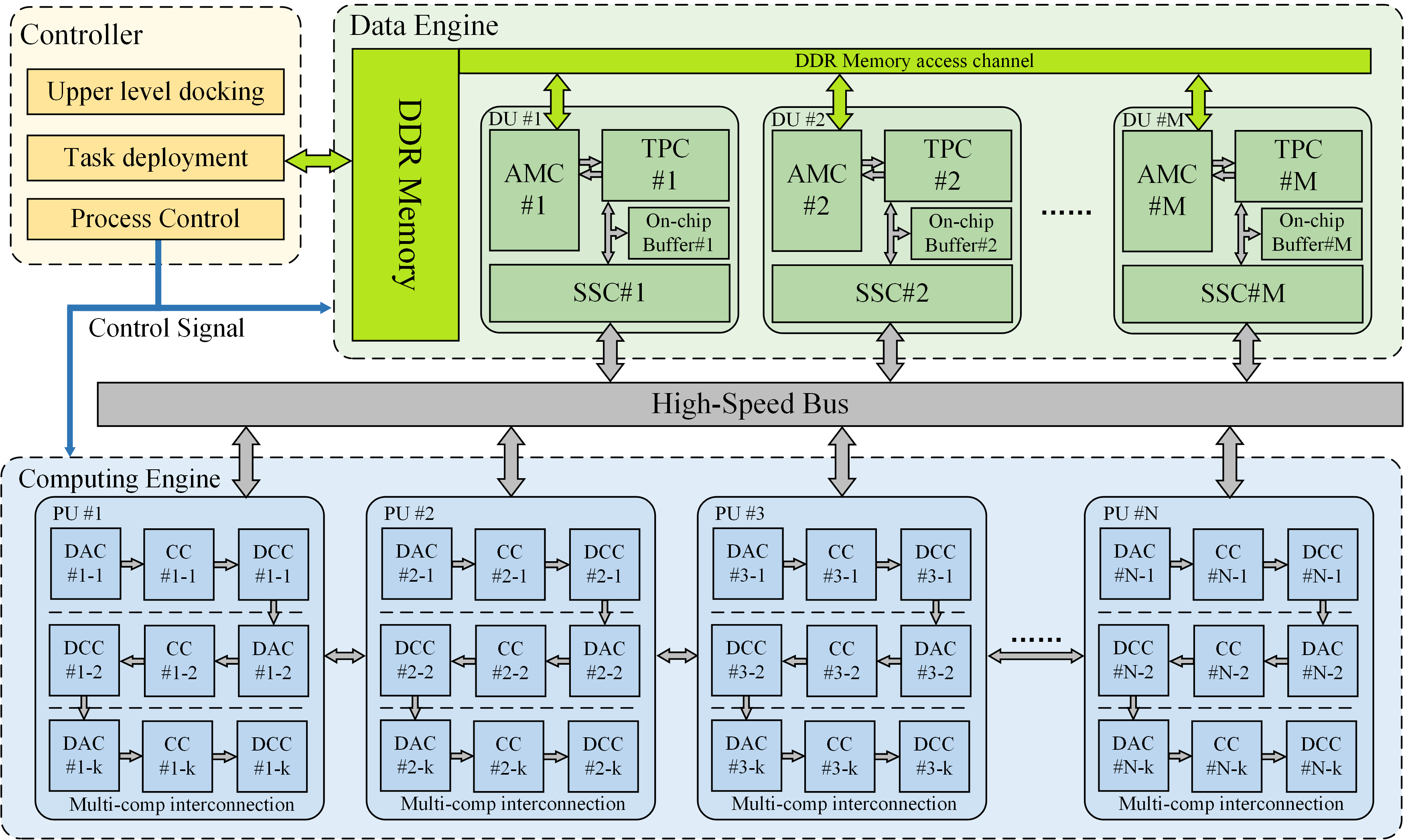}
  \caption{EA4RCA framework architecture.}
\end{figure}

\begin{table*}[t]
  \caption{Names and Functions of Components in the EA4RCA Framework.}
  \newcommand{\tabincell}[2]{\begin{tabular}{@{}#1@{}}#2\end{tabular}}
  \label{tab:commands}
  \begin{tabular}{ccl}
    \toprule
    Abbreviation &Full Name & Function\\
    \midrule
    AMC	&Memory Access Component	&Provide DDR Memory access services. \\
    TPC	&Task Processing Component	&Splitting and summarizing tasks.\\
    SSC	&Stream Service Component	& \tabincell{l}{Communicate with computing engines;\\send and receive task data.}\\
    DAC	&Data Allocation Component	&Assign data to multiple cores in CC.\\
    CC	&Computing Component	&Calculation and processing of tasks.\\
    DCC	&Data Collection Component	&Collect results from multiple cores in CC.\\
    \bottomrule
  \end{tabular}
\end{table*}

After RCA applications are divided into regular computing and communication parts, there will be hardware requirements for high-speed computing and high-speed data services. Considering that different hardware architectures are good at different types of tasks, the framework includes three parts: computing engine, data engine, and controller. The three parts of computing, data, and control are deployed and implemented on the hardware architecture that is good at, which can improve efficiency and make full use of hardware resources. The EA4RCA framework structure is shown in Figure 1. The EA4RCA framework first designs the AIE according to the communication rules of the application, designs the data stream and off chip data supply with the AIE data requirements as the core, and adds parallel processing, so that the framework can have strong performance performance in low communication class and high communication class RCA applications.

The EA4RCA framework employs a "top-down" design approach that is guided by application characteristics. This approach allows developers to rapidly obtain solutions that are relatively superior and feasible. It greatly aids in industry development. However, one drawback of this design approach is that the solutions obtained may not necessarily be optimal. Nonetheless, we will delve into this matter in our subsequent research. To facilitate broader application compatibility, it further decomposes the computation engine, data engine, and controller into multiple abstract components and defines their functionalities. During the actual deployment of applications, multiple implementation options can be provided for each abstract component. Depending on the specific requirements and application characteristics, the appropriate implementation option can be selected, enabling component replacement and updates without affecting other parts. This ensures the framework's strong adaptability. The names and functionalities of the components in Figure 1 are presented in Table 1, and the roles and characteristics of each component will be described in detail later in the paper.

The computing engine serves as the core of the framework and is responsible for efficient task solving. It consists of multiple sets of processing units (PUs) that can solve the decomposed sub-tasks. The implementation logic of each PU can be the same or different. Having the same implementation for PUs can increase the parallelism for processing tasks of the same type, while different PU implementations can handle different types of tasks. In the execution flow, there are also data channels between PUs for data exchange, but they are only open during communication phases.

The data engine serves the computing engine and is responsible for task decomposition and data flow services. It requires strong data manipulation capabilities and high-speed on-chip caching. The data engine includes DDR memory and multiple sets of data units (DUs). The DDR memory is used to store task data deployed by the controller, while the DUs handle task decomposition, aggregation, and interaction with the computing engine. In the EA4RCA framework, a DU typically serves multiple PUs, and this coordination is referred to as DU-PU pairs. The framework includes multiple DU-PU pairs executing in parallel.

The controller is responsible for upper-level integration, task deployment, and flow control. It first receives specified tasks from the upper-level and then synchronizes task data to the data engine for task deployment. Finally, it controls the flow of the framework's operation, working in collaboration with other components to complete the computations.

\subsection{EA4RCA framework execution modes}

RCA type algorithms themselves have regular computing and communication characteristics, and are easy to decouple. Therefore, the EA4RCA framework splits computing and communication in RCA type applications, and adopts the design mode of regular communication to reduce the number of times that computing is interrupted, so as to increase the proportion of computing time in AIE runtime. This is because when developing accelerators based on the ACAP hardware architecture, we found that separating the computation and communication of AIEs, making them aggregate operations, and making full use of the DMA engine inside AIEs can significantly improve performance. At the same time, this splitting method has also been proved to be extremely effective in CA\cite{CA1} and DAE\cite{DAE}. The AIE core has two communication modes: Stream (1.95TB/s) and DMA (15.6TB/s), Stream can communicate at the core runtime, DMA can only move large pieces of data when the core is turned off. The experiment shows that when using the AIE single core to calculate 32x32x32 matrix multiplication under the ideal simulation state, three ways of (1) AIE Stream+communication calculation crossover, (2) AIE Stream+communication calculation aggregation, and (3) AIE DMA+communication calculation aggregation are tested respectively. The test results are shown in Table 2.

\begin{table*}[t]
  \caption{Simulation test results of three methods.}
  \newcommand{\tabincell}[2]{\begin{tabular}{@{}#1@{}}#2\end{tabular}}
  \label{tab:commands}
  \begin{tabular}{ccccc}
    \toprule
    Method&Data Type&Communication size&Overall FLOP&Run time\\
    \midrule
    (1) AIE Stream + Crossover&Float&16&65536&31.06us \\
    (2) AIE Stream + Aggregation&Float&1024&65536&8.61us\\
    (3) AIE DMA + Aggregation&Float&1024&65536&3.49 us\\
    \bottomrule
  \end{tabular}
\end{table*}

In method (1), since the core needs to continuously receive data from the Stream to maintain the calculation, the calculation is constantly interrupted, resulting in a long running time. In method (2), the data required for multiple calculations is first received from the Stream at one time and then calculated in the AIE memory. This avoids AIE operations being interrupted, and has a higher data reuse rate. Compared with the previous mode, the performance is enhanced. In method (3), with the help of the high-speed data handling of the AIE DMA engine, the communication time is greatly shortened, and AIE can also focus on computing, so the performance is the highest. It can be seen that under the same FLOP number, this way of separating the calculation and communication of AIE can improve the overall performance of the system.

Although the separation of computing and communication has greatly improved the performance of AIE, the on-chip storage overhead caused by communication convergence has increased, because the data transmission during communication is still stored on the chip, and when the amount of one-time communication data increases to a certain extent, AIE performance will no longer be enhanced. Therefore, when designing AIE, it is necessary to determine the amount of data to be communicated at a time by combining the storage space provided by the hardware itself and the application characteristics, so as to balance the on-chip storage pressure and AIE computing efficiency as far as possible.

The EA4RCA framework consists of multiple DU-PUs pairs that can operate in parallel. Each DU-PUs pair's execution flow is divided into two major stages: the computation stage and the communication stage. These stages are executed in an alternating manner, corresponding to the computational and communication requirements of the divided RCA application. The execution flows of individual DU-PUs pairs are independent of each other, as illustrated in Figure 2. From a holistic perspective of the framework, these DU-PUs pairs may be in different stages simultaneously. They can collaborate to form a pipelined execution state or independently complete different tasks.

\begin{figure}[t]
  \centering
  \includegraphics[width=0.8\linewidth]{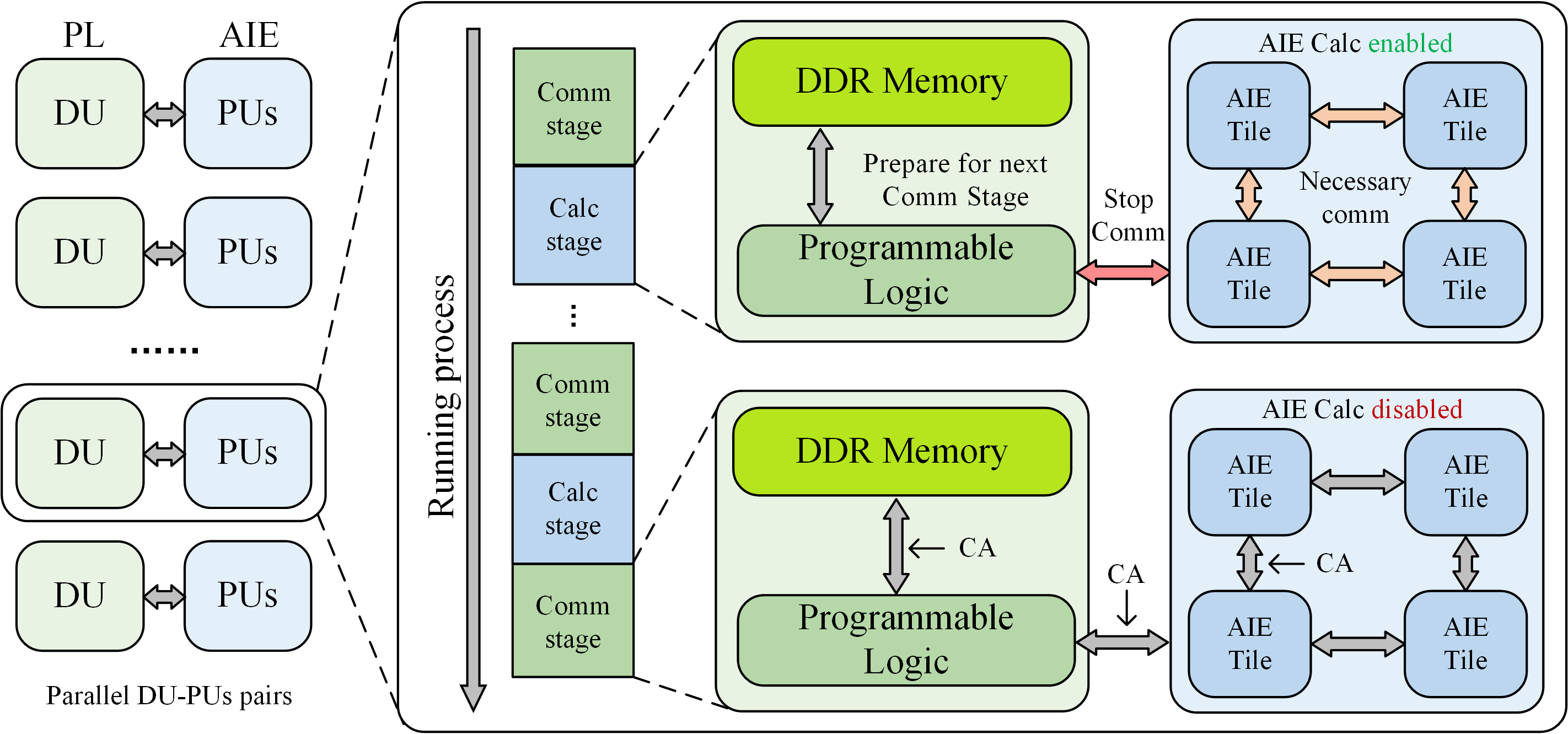}
  \caption{EA4RCA framework running process.}
\end{figure}

During the computation phase, the AIE computation enable signal in the PU is activated, allowing it to focus on computations. Communication between the DU and PUs is halted, and only internal communication necessary for the computation is maintained within the AIE. Simultaneously, the DU prepares data for the next communication phase.

During the communication phase, the AIE computation enable signal is turned off, and the communication channels between different hardware components are fully opened. Communication is guided by CA optimization, aiming to minimize the time overhead caused by communication interruptions during system operation. The EA4RCA framework is dedicated to reducing the time spent on communication.

Accelerating both phases can enhance the overall workflow. Increasing the number of parallel computing cores can reduce the time required for the computation phase, thereby improving the overall speed. Additionally, the CA concept is utilized to optimize AIE inter-core communication and AIE off-chip communication, aiming to reduce the time consumed during the communication phase.

In more complex applications, dividing the workflow into multiple stages may lead to improved performance. However, this approach also increases design complexity and introduces additional challenges. In the future, we will explore more complex RCA application scenarios and make new attempts.This paper focuses on relatively simple RCA applications within the EA4RCA framework, where dividing the workflow into two stages has already proven to be sufficient for achieving the desired outcomes.

However, in non RCA applications with irregular communication and computation, non RCA applications are characterized by unfixed communication times and frequencies, and large traffic. However, this framework can still be used for design, which will lead to performance degradation. In non RCA applications, the AIE Stream can be used to dynamically send and receive data when the AIE core is running. At this time, the method of data filling and AIE memory buffering and multiplexing (method (2)) can be used to achieve a certain degree of separation between communication and computing, so as to improve the efficiency of AIE as much as possible.

\subsection{Computing Engine}

The computing engine is implemented based on AIE and consists of multiple processing units (PUs). The structure of a PU is illustrated in the left part of Figure 3. The PU is composed of a multi-level processing structure, which includes the Data Allocation Component (DAC), Computing Component (CC), and Data Collection Component (DCC). The functions of these three components are described in Table 1.

The AIE plays a central role in computation, and all computational tasks are allocated to the AIE cores. However, due to the large number of AIE cores in the array, it is necessary to organize as many cores as possible to run concurrently during computation. These cores form the CC. Additionally, these cores need to be assigned data and collect their computational results, which requires the presence of the DAC and DCC components.

During the processing in a PU, a subtask may consist of multiple processing stages. A Processing Structure (PST) is responsible for addressing one of the processing stages within the subtask. If a subtask has only one processing stage, the PU consists of a single processing structure.

\begin{figure}[t]
  \centering
  \includegraphics[width=0.9\linewidth]{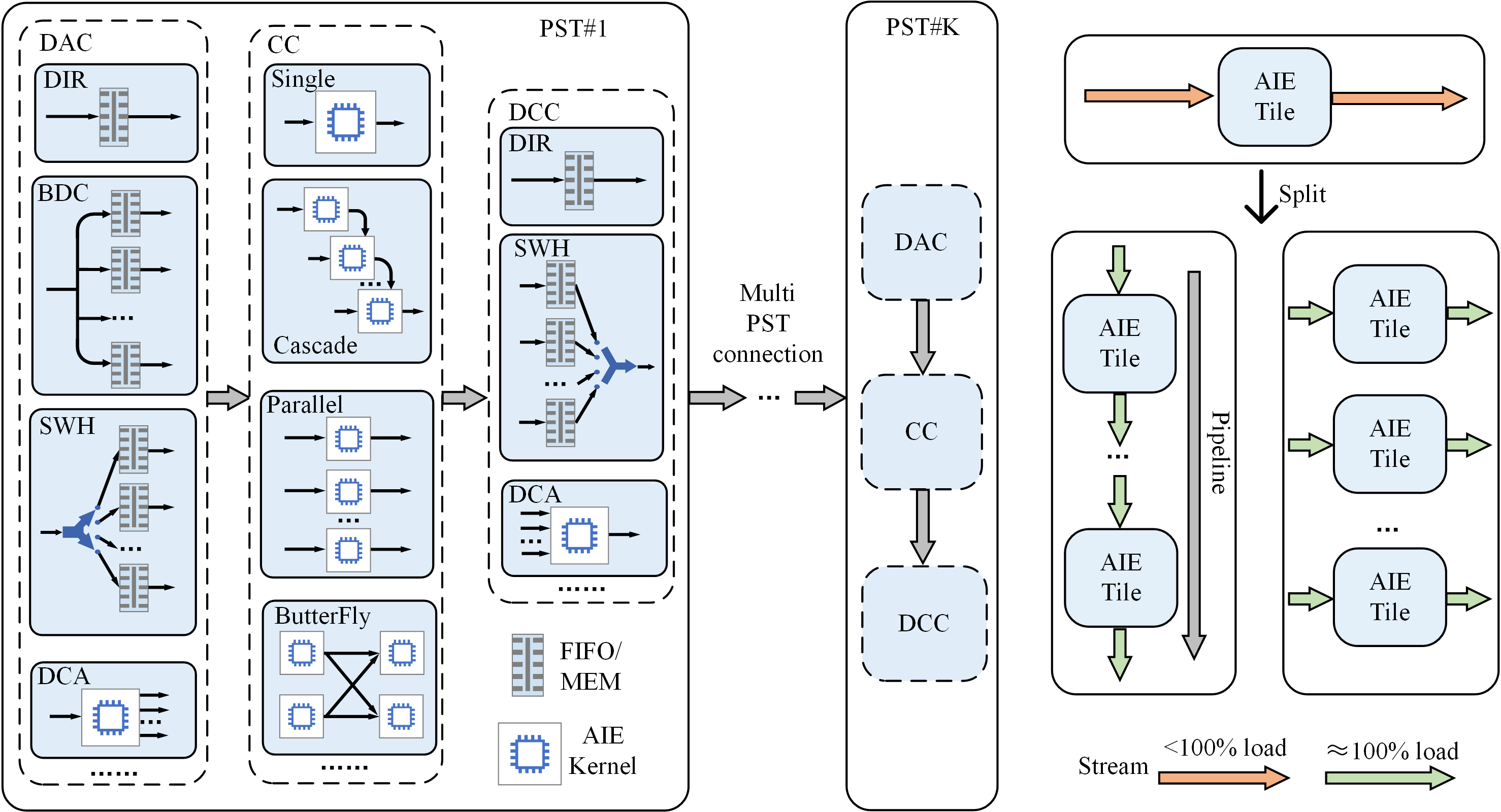}
  \caption{Processing unit architecture.}
\end{figure}

1) Computing component

Each iteration of the PU can solve one subtask, with the CC being the most crucial component. The CC is activated during the computation phase. Due to the flexibility of AIE programming, a single core in the CC can fulfill the computational requirements. However, in most cases, the computational speed of a single core in the CC may not match the data flow rate, resulting in the DU waiting for the PU and causing idle time.

To minimize the waiting time for each hardware component, the computational capacity of the PU should be similar to the data processing capability of the DU. Therefore, it is necessary to further divide the subtask into multiple cores to form a multi-core CC, as shown in the right part of Figure 3. This enables the CC to match the data flow rate. Additionally, the multiple cores should be balanced in terms of workload distribution, optimizing the overall performance.

According to the calculation characteristics of different applications, different implementation modes of CC can be written. We give four common implementation modes of CC. In other applications, if these modes do not match the calculation characteristics of the application, new implementation modes can be added to the calculation unit, or the original modes can be combined, and finally suitable DAC and DCC can be equipped for it.

\begin{itemize}
\item {\verb|Single|}: In the case where the computational workload is relatively small and the processing speed of a single core is sufficient to match the data processing rate of the DU, a single-core processing approach is suitable.
\item {\verb|Cascade|}: Multiple cores can be interconnected to form a pipeline with the objective of distributing the computational workload from a single core to multiple cores, thereby accelerating the computation while maintaining the same functionality as the Single approach. In this configuration, each core handles a portion of the computation for a specific sub-problem. The accumulators are passed down through each stage of the pipeline, and the final computed result is generated by the last-stage core.
\item {\verb|Parallel|}: In scenarios where different types of computational operations are required for processing sub-problems, multiple parallel and non-interconnected single-core or Cascade-mode core groups are utilized. This approach is commonly used in conjunction with the Cascade mode. Each core group is responsible for handling a specific type of computation, allowing for efficient processing of diverse computational operations within the sub-problems.
\item {\verb|Butterfly|}: An operational component specially equipped for butterfly operations.
\end{itemize}

2) Data allocation  and collection component

DAC and DCC are responsible for distributing data to individual cores and collecting results in CC. During the computation phase of CC, DAC and DCC ensure the necessary communication for the computation. During the communication phase, data exchange occurs among DACs and DCCs in each processing unit (PU). To accommodate the diverse data requirements of different applications, the framework provides the following four implementations for DAC:

\begin{itemize}
\item {\verb|Direct(DIR)|}: Directly connects the output data stream of the data engine to the computing component which is only applicable to the single core computing component.
\item {\verb|Broadcast(BDC)|}: Copies the output of the data engine and sends it to the specified core of the computing component within one cycle.
\item {\verb|Switch(SWH)|}: Time-sharing takes up a channel to send data to different computing component cores, usually when the computing speed of a single core is lower than the data distribution speed and there are more cores in need of service.
\item {\verb|Dedicated Core Allocation(DCA)|}: Specially equipped with a core for logically complex data organization or distribution, which allows for the existence of computational operations internally.
\end{itemize}

For DCC, its structure and characteristics are generally similar to DAC as they both serve the CC and are used to collect the output results of CC. However, since broadcasting is not applicable during data collection, the framework provides three implementations for DCC: DIR, SWH, and DCA. Additionally, new implementations can be added to both DAC and DCC.

In practical applications, a CC system may employ multiple data allocation strategies or require multiple batches of data within a single iteration of PUs. This means that multiple different types of DAC and DCC can be connected to a CC, and they can serve only a subset of AIE cores within the CC. Therefore, the framework allows the presence of multiple DACs and DCCs in a processing structure.For applications that require communication between PUs, the output of DAC and the input of DCC can still be connected to other PUs. However, the data channels between PUs are only open during the communication phase. To reduce communication overhead, it is advisable to minimize inter-PU communication as much as possible when designing the application deployment scheme.

When the application program has acceleration requirements of different operators, because each AIE is fully customizable, multiple PUs of different types can be designed to provide acceleration services of different operators. At the same time, in order to further accelerate the running speed, multiple PUs of the same type can be deployed under the condition of hardware resources to increase the parallelism of processing. Finally, the data engine can schedule these PUs to run orderly.

\subsection{Data Engine}

The data engine is responsible for fulfilling the data input and output requirements of the AIE. AMD/Xilinx officially provides a data mover in the PL (Programmable Logic) domain as a data engine serving the AIE\cite{vitis_lib}. However, this approach has the following shortcomings:

\begin{itemize}
\item It only supports sequential access to the onboard DDR memory, lacking flexibility.
\item Each DDR port can only serve one PLIO (Programmable Logic Input/Output) path to the AIE, limiting the ability to serve multiple paths simultaneously. Additionally, the maximum rate of PLIO is 128b/cycle, while each DDR port can provide a maximum rate of 512b/cycle, resulting in hardware resource waste.
\item It does not provide task decomposition and aggregation functionality, often requiring additional pre-processing in the PS (Processing System) or within the AIE, which hinders overall system performance improvement.
\end{itemize}

Furthermore, there is currently a lack of comprehensive research on high-performance data engines with complete functionality. Therefore, we propose a high-performance data engine based on the PL domain, which significantly enhances functionality and flexibility while ensuring performance.

The data engine of the EA4RCA framework enables efficient task decomposition and aggregation, catering to the data exchange requirements of each processing unit (PU) in the computing engine. During the communication phase, the data engine communicates with the computing engine, while during the computation phase, it prepares data for the next round of communication phase. The data engine consists of DDR memory and multiple data units (DUs). Each DU comprises three components: Memory Access Component (AMC), Task Processing Component (TPC), and Stream Service Component (SSC). The functionalities of these components are depicted in Table 1. These components execute in parallel within the programmable logic (PL), interconnected using internal data streams. Furthermore, each DU's components can be implemented using different approaches. The framework provides predefined implementations for these three abstract components and also allows developers to add new implementation methods.

\begin{algorithm}[t]
    \caption{AMC Reader Algorithm}
    \label{alg:AOA}
    \renewcommand{\algorithmicrequire}{\textbf{Input:}}
    \renewcommand{\algorithmicensure}{\textbf{Output:}}
    \begin{algorithmic}[1]
        \REQUIRE $MemoryBlock$, $MemorySize$, $AddrSeq$, $BurstSize$, $N$, $Mode$  %%input
        \ENSURE $hls::stream$    %%output
        
        \STATE  $int Addr \leftarrow0; //$ The actual memory location currently being accessed.
        \STATE  $ap\_uint \, tmp \leftarrow NULL;//$ The data read out.
        \IF {$Mode = CSB$}
            \STATE \textbf{//HLS compiler automatically infers the number of FOR loop to execute burst memory access.}
            \FOR{each $i \in [1,MemorySize]$}
                \STATE $tmp \leftarrow MemoryBlock[i];$
                \STATE $hls::stream.write(tmp);$
            \ENDFOR
        \ENDIF
        \IF {$Mode = JUB$}
             \FOR{each $n \in [1,N]$}
                \STATE $Addr \leftarrow AddrSeq.read();$
                \STATE \textbf{//This FOR loop performs burst memory access.}
                \FOR{each $i \in [1,BurstSize]$}
                    \STATE $tmp \leftarrow MemoryBlock[Addr+idx];$
                    \STATE $hls::stream.write(tmp);$
                \ENDFOR
            \ENDFOR
        \ENDIF

         \IF {$Mode = UNOD$}
             \FOR{each $n \in [1,N]$} 
                \STATE \textbf{//Addr cannot be predicted, burst memory access does not occur here.}
                \STATE $Addr \leftarrow AddrSeq.read();$
                \STATE $tmp \leftarrow MemoryBlock[Addr];$
                \STATE $hls::stream.write(tmp);$
            \ENDFOR
        \ENDIF
    \end{algorithmic}
\end{algorithm}

1)Access memory component

The AMC utilizes the M\_AXI bus to perform read and write operations on the DDR memory, and its performance directly affects the overall system performance. In practical applications, it is desirable to make full use of the memory bandwidth by leveraging burst transfers through the bus. However, certain applications may have requirements for non-sequential or unordered memory access, which do not meet the conditions for burst transfers. To meet various memory access demands and achieve high memory access performance under different requirements, the framework provides three implementations for the memory access component.

\begin{itemize}
\item {\verb|Complete Sequence Burst(CSB)|}: Read or write data sequentially in memory address order to maximize memory access performance.
\item {\verb|Jump Burst(JUB)|}: Use a different start address to continuously access a piece of data in memory after this address and use bus burst mode, performance is inferior to CSB mode.
\item {\verb|Unordered(UNOD)|}:The storage access address cannot be predicted, and the bus burst mode cannot be used. The performance is the worst, but the advantage is high flexibility.
\end{itemize}

The logic for memory read operations in the AMC is illustrated in Algorithm 1. The logic for memory write operations is similar to memory read operations. The AMC requires the following parameters: MemoryBlock, MemorySize, AddrSeq, BurstSize, N, and Mode. These parameters represent the memory pointer, the number of data elements in the memory, the memory address sequence, the number of data elements accessed in each burst, the execution count, and the AMC mode, respectively.

In the CSB mode, the AMC sequentially accesses the MemoryBlock from the beginning to the end based on the MemorySize parameter. No additional parameters are required for this mode.

In the JUB mode, the first step is to read the starting memory address from AddrSeq. Then, based on the BurstSize parameter, the AMC performs burst memory accesses. When burst accesses are used, the HLS compiler automatically infers and utilizes burst transfers based on the inner loop iterations (for each i).

In the UNOD mode, due to the unpredictable memory address sequence, burst transfers cannot be used, the AMC can only store or retrieve a single data element from memory based on the address provided by AddrSeq.

2)Task processing component

\begin{figure}[t]
  \centering
  \includegraphics[width=0.8\linewidth]{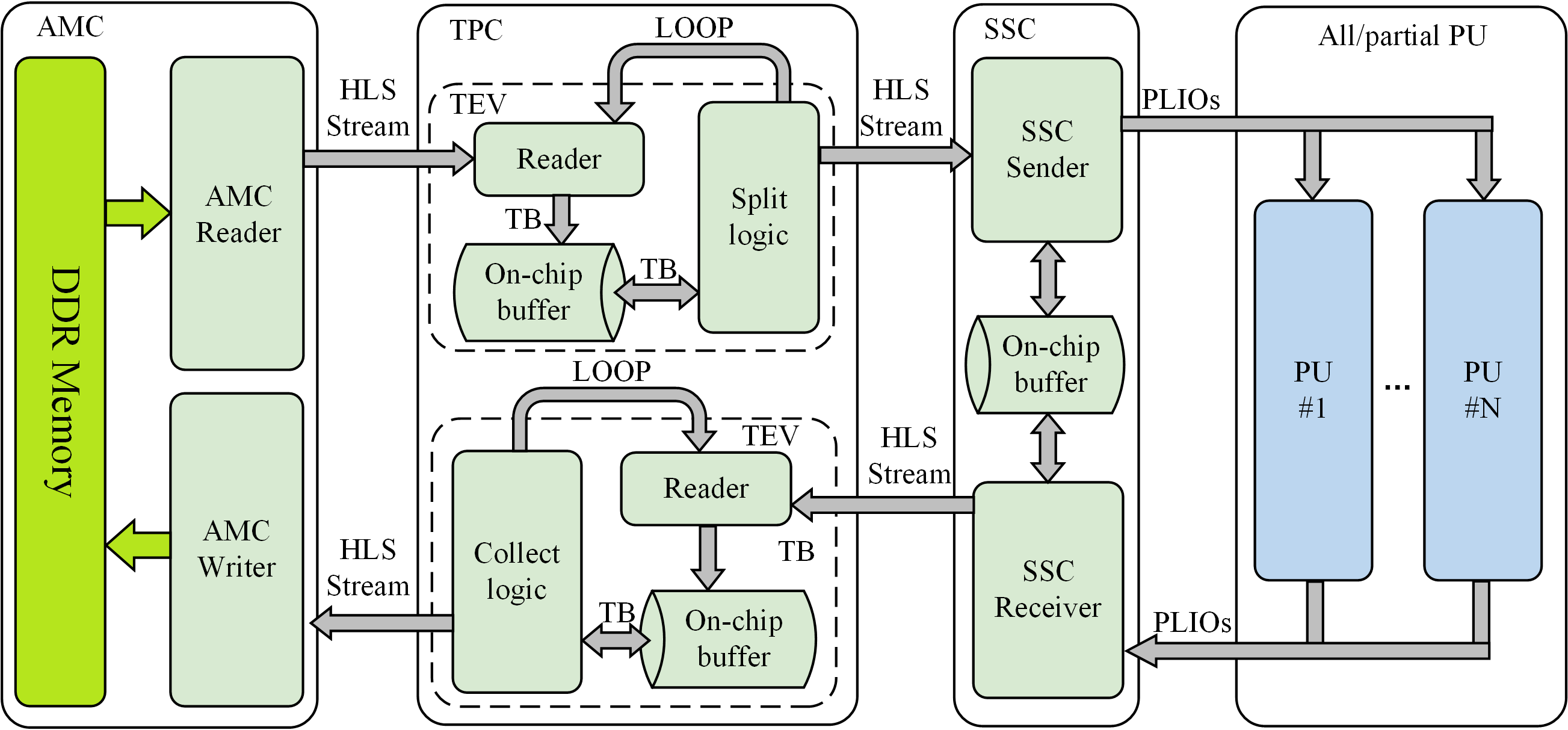}
  \caption{Task processing component structure.}
\end{figure}

The Task Processing Component (TPC) is the core part of the DU and is responsible for task decomposition and aggregation of computational results. The structure of the TPC is depicted in Figure 4. The AMC and SSC components isolate the TPC from external interference, allowing the TPC to work without disruptions. The specific task decomposition and aggregation logic of the TPC will depend on the particular application.Furthermore, the presence of the TPC enables the division of tasks of different scales into multiple fixed-size subtasks. This capability allows various applications deployed on this framework to naturally support adaptive characteristics.

The process of decomposing or aggregating a task in the TPC is referred to as a Task Event (TEV). The execution of a TEV can consume a Task Block (TB) from either the AMC or the computing engine. The TB is stored in an on-chip cache, and it represents the minimum data set required for a TEV. During the execution of a TEV, the necessary TB is first fetched from the AMC or SSC into the on-chip cache. Then, the decomposition or aggregation logic is applied to process the TB into the appropriate format. Finally, the resulting data is either output to the SSC or written back to the AMC.

In a single iteration of a PU, the total amount of data to be transferred may involve multiple TBs. This means that multiple TEVs are required to complete one iteration of the PU. This approach accommodates situations where the data transfer during PU processing is not fixed or when the sub-problem size is too large to be transferred in a single operation.

To address the specific characteristics of task decomposition in different applications, the framework provides three implementation options for the Task Processing Component.

\begin{itemize}
\item {\verb|Cache Update (CUP)|}: Execute multiple TEV, each TEV reads a new batch of TB from AMU or SSC, updates the buffer, processes it and sends split or summary results.
\item {\verb|Cache Hold(CHL)|}: executes TEV for a specified number of times and keeps TB from being updated in the buffer. CHL is used when the total amount of data is small but the computation is heavy, or when only fixed tasks need to be processed repeatedly.
\item {\verb|Through(THR)|}: Suitable for applications that do not require any disassembly of tasks, do not have TEV, and directly connect MAU output to SSC input without on-chip buffer.
\end{itemize}

3) Stream service component

While the TPC is responsible for task decomposition and aggregation, it lacks the knowledge of how to correctly map multiple tasks to their corresponding PUs. This is where the SSC comes into play. The SSC is responsible for handling the interaction between the computing engine and the TPC. It consists of a Sender and a Receiver component. The SSC ensures the correct functionality while enabling tasks to be transferred between the computing engine and the data engine at a high rate.

Additionally, the SSC can utilize the on-chip cache within the DU. This allows the SSC to parallelize the transmission and reception of data to and from multiple PUs, without conflicting with the buffer access of the TPC.

\begin{figure}[t]
  \centering
  \includegraphics[width=0.9\linewidth]{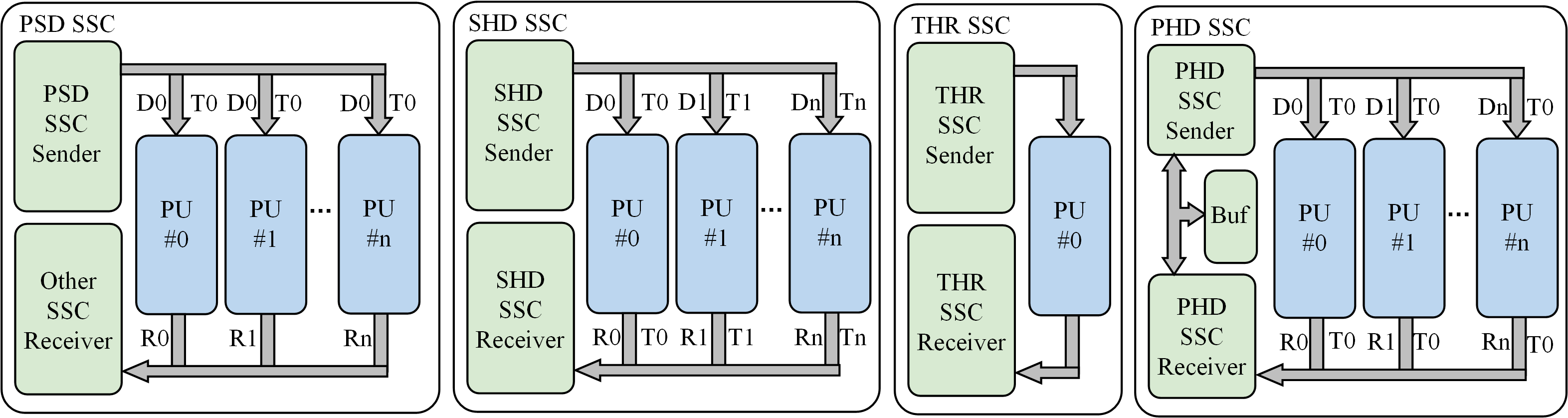}
  \caption{SSC four service mode structure and service timing.}
\end{figure}

In various applications, considering the varying number of PUs and their different processing speeds, as well as the diverse data requirements for each PU, the framework in SSC provides four implementation approaches for the Sender and Receiver. The service timing for the four SSC approaches is illustrated in Figure 5.

\begin{itemize}
\item {\verb|Parallel Same Data(PSD)|}: When multiple PUs require the same data, the same subproblem is sent to multiple PUs in parallel. This method is only for the Sender.
\item {\verb|Serial Heterogeneous Data(SHD)|}: Sends different subproblems to or receives different subproblem results from PU to serve each PU in a serial manner.
\item {\verb|Parallel Heterogeneous Data(PHD)|}: The effect is the same as SHD. The difference is that various PUs are served in a parallel way. However, all subproblem data to be sent needs to be read into the buffer before parallel transmission.
\item {\verb|Through(THR)|}: Connects TP directly to PU and only one PU can be served.
\end{itemize}

Although the SHD and PHD approaches achieve the same outcome, there is a distinction in how they handle PU service. In the SHD approach, if one or more PUs have slower processing speeds, the SHD method will wait until those PUs finish their service before proceeding to serve the subsequent PUs. This waiting time can have an impact on overall efficiency. On the other hand, the PHD approach does not encounter this issue but requires a certain amount of on-chip storage space.

\subsection{AIE Graph Code Generator}

To enhance the optimization and standardization of the AIE development process, we introduce the AIE Graph Code Generator.  Users can generate the compileable AIE engineering code of the PU in the calculation engine by one click through the Graphical User Interface (GUI) or importing the configuration file, greatly improving the efficiency of AIE development and reducing the difficulty of development.   In the experimental chapters, the PU in the calculation engine is generated by the AIE Graph Code Generator.   The AIE Graph Code Generator Structure is shown in Figure 6.   The GUI PU Editor Code repository, generator core and Xilinx backend.

\begin{figure}[t]
  \centering
  \includegraphics[width=\linewidth]{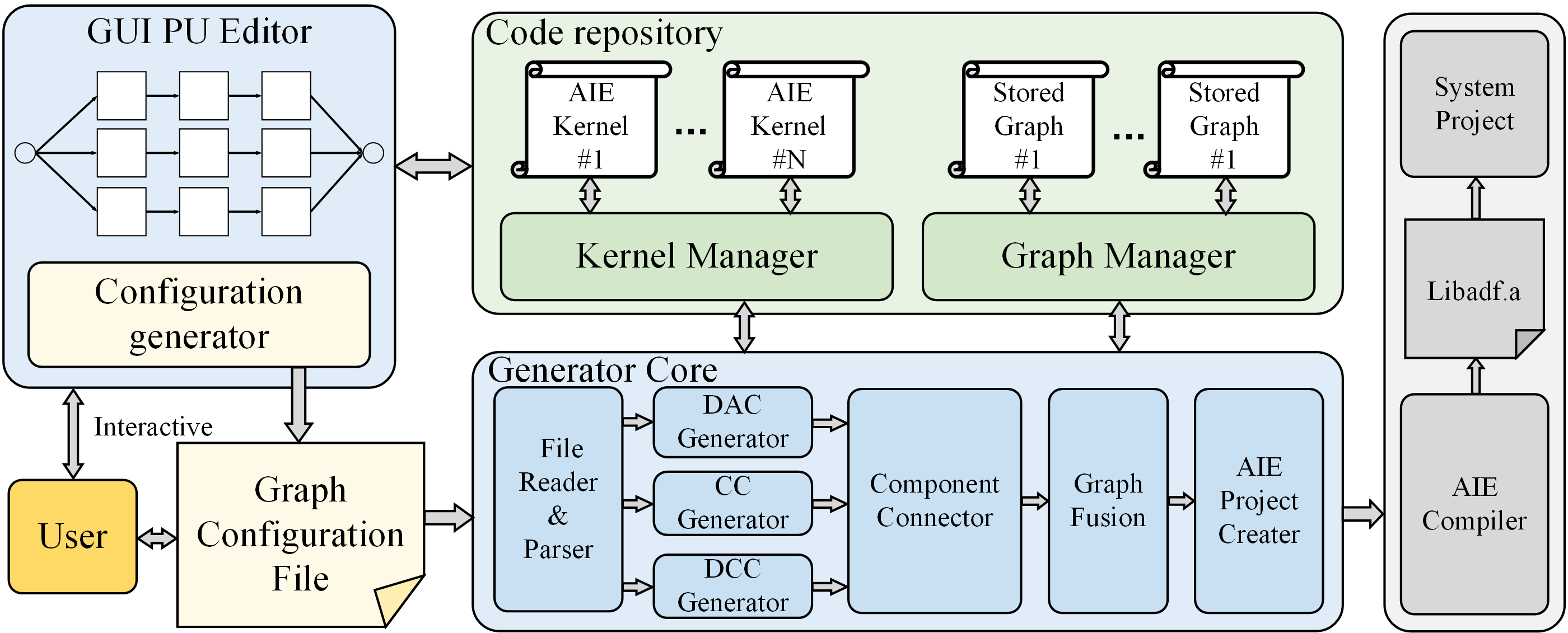}
  \caption{AIE Graph Code Generator Structure.}
\end{figure}

The Generator Core is the core component of the AIE Graph Code Generator, and compileable AIE project code is generated through the imported Graph Configuration File. The Generator Core first parses the PU information of the configuration file, which contains DAC The detailed configuration of the connection between CC and DCC itself, through the DAC generator CC Generator and DCC Generator generate instances of the three parts, and then copy, sort and connect the instances of the three parts through Component Connector to form a complete PU structure, Graph Fusion is responsible for integrating the saved Graph structure into the current design (optional), and finally using AIE Project Creator to generate AIE project code for compilation.

The Code Repository manages the AIE kernel source code and the Stored Graph structure through the Kernel Manager and Graph Manager, When the AIE kernel source code is used for GUI interaction and generates code with the Generator Core, Stored Graphs are complete designs that have been saved previously. They can be integrated into the current design or used for storing backups.

GUI PU Editor is mainly responsible for interacting with users. Users can freely create and edit DAC, CC, DCC using graphical interface, users first need to create CC containing AIE kernel and allocate kernel source code for it, then can freely combine according to the communication mode supported by AIE hardware to form DAC and DCC, and finally generate connectable nodes to connect with PLIO or AIE kernel. Since the Generator Core needs Graph Configuration File input, the GUI PU Editor contains a Configuration generator to convert the user's editing results into configuration files. In addition, users can also directly edit the Graph Configuration File to achieve the same effect.

Xilinx backend includes a tool chain that can compile AIE projects. It can compile, simulate, and analyze AIE code, The AIE Graph Code Generator integrates the AIE design into the complete ACAP hardware design by compiling the libadf.a file using Xilinx backend.

\section{EVALUATION}

\subsection{Experimental setup}

The EA4RCA framework was implemented on the Xilinx VCK5000 Versal Development Card, which is based on the Xilinx 7 nm Versal ACAP architecture.   The software environment was constructed using the Xilinx Vitis 2022.2 toolchain, while the assessment of device power consumption was conducted using Power Design Manager (PDM) 2023.2.2. To thoroughly validate the effectiveness and adaptability of various RCA application accelerators based on the EA4RCA framework, we employed the EA4RCA framework to design accelerators for four RCA applications: matrix multiplication (MM), Filter2D, Fast Fourier Transformation (FFT), and Performance Testing of AIE Computing Based on MM (MM-T).   These applications entail regular computation and communication, making them amenable to effective decoupling.   Table 3 presents the problem sizes and data types used in the experiment. This chapter will describe these four accelerators in detail, showing how to carry out the overall design and customization to implement the accelerator and evaluate the performance.

\begin{table*}[t]
  \caption{Problem Size and data type.}
  \label{tab:commands}
  \begin{tabular}{cccc c}
    \toprule
     Item&	MM&	Flter2D&	FFT&	MM-T\\
    \midrule
    Problem Size& \makecell{768x768x768\\1536x1536x1536\\3072x3072x3072\\6144x6144x6144}& \makecell{128x128,5x5\\3480x2160,5x5\\7680x4320,5x5\\15360x8640,5x5}& \makecell{1024\\2048\\4096\\8192}&32x32x32\\
    Data Type&	Float&	Int32&	Cint16&	Float\\
    \bottomrule
  \end{tabular}
\end{table*}

\subsection{Accelerator design process based on EA4RCA framework}

We will show how to use the EA4RCA framework to design a high-performance MM accelerator to truly reflect the design process. MM accelerator is usually called as an operator, so it is necessary to consider the high performance of the overall system operation process, not only to pursue the extreme performance of one part and increase the cost of other parts, but also to consider the adaptability of the task scale, so that operators can be used more widely.

When using the EA4RCA framework to implement the accelerator, you should first consider the design of CC in the PU, that is, how to split large-scale tasks into multiple subtasks and map them to each AIE kernel in CC. In the MM accelerator, we choose the block division method of matrix parallelization operation, because this division method can divide the MM tasks of variable size into multiple MM tasks of fixed size, AIE only needs to concentrate on solving fixed size sub tasks,   and complete the overall task through multiple iterations, which is conducive to the design and implementation of AIE.

Therefore, we should determine the task size of each AIE.   In the previous study, Jason Cong's team from UCAL mentioned in CHARM\cite{acap_impl3} that when considering the AIE kernel efficiency and the AIE DMA capacity limit, the FLOAT type MM operation of 32 × 32 × 32 size is optimal, so we also use the same AIE single core load.   In this case, when calculating MM of M × K × N size, The total number of iterations required for the AIE single core can be calculated using Formula 1.

\begin{eqnarray}
\text { Iter }_{\text {Kernel }}=\left\lceil\frac{M}{32}\right\rceil \times\left\lceil\frac{K}{32}\right\rceil \times\left\lceil\frac{N}{32}\right\rceil
\end{eqnarray}

After determining the load of a single core, we can use one AIE core to complete MM operations without considering the time cost. However, in order to make full use of the 400 AIE cores, we need to build a PU containing multiple AIE cores. PU can organize multiple AIE cores to complete larger MM operations in one iteration. We specify that the size of the MM completed by the PU is 128 × 128 × 128, and the total number of iterations calculated by Formula 1 is 64. Therefore, the internal CC is organized as Parallel<16>* Cascade<4>. There are 64 AIE cores in total, and each Cascade<4>outputs a 32 × 32 result. Through 16 groups of Parallel Cascade<4>, the PU can complete 128 × 128 × 128 MM in one iteration.

After CC is determined, we need to determine the design of DAC and DCC, allocate and collect data for CC. In order to use fewer PLIOs, we choose to use SWH+BDC (packet switching+broadcasting) to improve the reuse rate of each PLIO data, SWH can distribute data for multiple cores in time without losing efficiency, BDC can broadcast data to multiple cores in one cycle. Finally, our DAC uses four PLIOs to send MatA, four PLIOs to send MatB, each PLIO sends four 32 × 32 matrices, and broadcasts them to the corresponding Cascade<4>according to the matrix multiplication rule, and the data of each PLIO is multiplexed four times. For the DAC, we use the PLIO of four SWH (packet switching) modes to collect the results of 16 channels of Cascade<4>.

\begin{figure}[t]
  \centering
  \includegraphics[width=\linewidth]{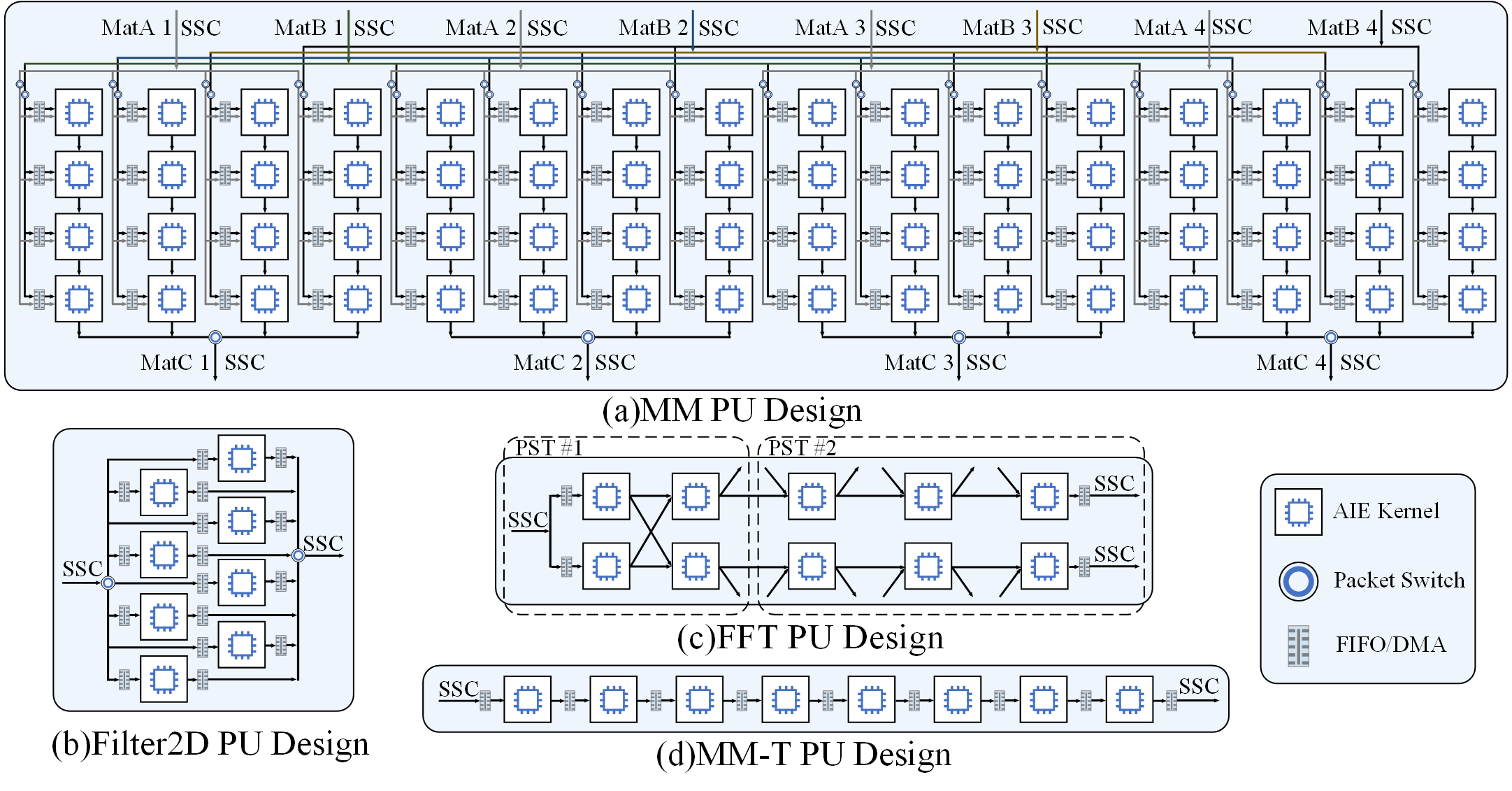}
  \caption{PU Design of Accelerators.}
\end{figure}

At this point, the MM PU design is complete, and the PU structure is shown in Figure 7(a). The one-time communication capacity of PU is three 128 × 128 × 128 matrices, and the AIE hardware resources consumed by each PU include 64 AIE cores and 12 PLIO ports. At this time, multiple copies of PU can be deployed in the AIE array if the hardware resources permit. We deployed six groups of such PU in the VCK5000 board to form the compute engine, which utilized a total of 384 AIE cores (96\%) and 72 PLIO ports. When using these resources to calculate MM of M×K×N size, the number of iterations required to calculate the engine can be calculated using Formula 2.

\begin{eqnarray}
\text { Iter }_{\text {Computing\_Engine }}=\left\lceil\left\lceil\frac{M}{128}\right\rceil \times\left\lceil\frac{K}{128}\right\rceil \times\left\lceil\frac{N}{128}\right\rceil {\div} 6 \right\rceil
\end{eqnarray}

Finally, we need to design a data engine for the computing engine. The data engine contains only one DU, The configuration between DU PUs pairs is 1:6, that is, one DU serves six PUs. The DU takes 27 128 × 128 matrices as TB when sending, AMC uses the JUB method to first read TB from DDR memory to the chip, accounting for 56\% of the URAM on the chip. These data can support nine iterations of the computing engine. Later, TPC splits TB into multiple 128 × 128 matrices, which are eventually sent by SSC to six PUs in parallel. When receiving, the DU uses six 128 × 128 matrices as TB, occupying 12\% of the URAM on the chip. Then it aggregates and accumulates through the TPC. Finally, the AMC writes the calculation results into the DDR memory. When DU-PUs pair is in communication phase, DU sends new data and receives calculation results, AIE suspends operation; During the calculation phase, PU focuses on calculation and does not communicate with DU. Meanwhile, DU starts to prepare the data required for the next iteration.

\begin{table*}[t]
  \caption{Implementation methods of components selected for applications.}
  \newcommand{\tabincell}[2]{\begin{tabular}{@{}#1@{}}#2\end{tabular}}
  \label{tab:commands}
  \begin{tabular}{cccc cccc}
    \toprule
     \multicolumn{1}{c}{\multirow{2}{*}{APP}}  & \multicolumn{4}{c}{Computing Engine} & \multicolumn{3}{c}{Data Engine} \\ 
       & PST  & DAC & CC & DCC  &AMC &TPC &SSC \\
    \midrule
    MM&	\#1&	SWH+BDC&Parallel<16>*Cascade<4>&SWH&JUB&CUP&PHD\\
    Filter2D&	\#1&	SWH&	Parallel<8>&	SWH&	JUB&	CUP&	PHD\\
    \multirow{2}{*}{FFT}	& \#1 & BDC & Butterfly & DIR &  \multirow{2}{*}{CSB} &  \multirow{2}{*}{CUP} &  \multirow{2}{*}{PHD} \\
    ~	& \#2 & DIR & Parallel<2>*Cascade<3>& DIR & ~ & ~ & ~ \\
    MM-T&	\#1&	DIR&	Cascade<8>&	DIR & Null	&	CHL&	THR\\
    \bottomrule
  \end{tabular}
\end{table*}

\begin{table*}[t]
  \caption{Hardware resource utilization.}
  \label{tab:commands}
  \begin{tabular}{cccc cccc c}
    \toprule
     Apps&	LUT&	FF&	BRAM&	URAM&	DSP&	AIE&	DU&	PU\\
    \midrule
    MM&	11403(7\%)&	105609(6\%)&	778(80\%)&	315(68\%)&	0(0\%)	&384(96\%)&	1&	6\\
    Filter2D&	248546(28\%)&	455277(25\%)&	526(54\%)&	0(0\%)	&168(9\%)&	352(88\%)&	11&	44\\
    FFT&	122650(13\%)&	214782(11\%)&	562(58\%)&	0(0\%)&	96(5\%)&	80(20\%)&	8&	8\\
    MM-T&	61039(7\%)&	96791(5\%)&	34(4\%)&	0(0\%)&	0(0\%)&	400(100\%)	&50&	50\\
    \bottomrule
  \end{tabular}
\end{table*}

At this point, the MM accelerators designed based on the EA4RCA framework are designed, and we discuss the performance of each accelerator in the following subsections. In this section, the MM accelerator is used as the main description object to show how to carry out the overall design and customization to implement the accelerator, and the design flow of the remaining three accelerators is consistent with the MM accelerator. The implementation of components selected for the four accelerators, MM, Filter2D, FFT and MM-T, is shown in Table 4, the hardware resources and DU-PUs pairs configurations are shown in Table 5, and the PU structure is shown in Figure 7.

\subsection{Performance analysis}

Based on the VCK5000 platform, we used the EA4RCA framework to deploy MM, Filter2D, FFT and MM-T, and conducted performance tests under different task scales and PU quantities. The hardware resource consumption and resource utilization are shown in Table 5. In order to comprehensively evaluate the actual performance of the hardware, we use Tasks/sec(TPS) and Giga Operations Per Second(GOPS) as speed indicators, and Power(W) and GOPS/W as energy efficiency indicators. To represent the actual performance of the system in actual hardware operation, these tests were run at a frequency of 1.33GHZ AIE and 300MHZ PL.

\begin{table*}[t]
  \caption{Performance of the MM accelerator at different task scales and PU quantities.}
  \label{tab:commands}
  \scalebox{0.9}{ %set scale
  \begin{tabular}{cccc cccc c}
    \toprule
    Problem Size&\makecell{Data\\Type}& \makecell{PU\\Quantity}&\makecell{Time\\(ms)}&Tasks/sec&GOPS&GOPS/AIE&\makecell{Power\\(W)}&GOPS/W\\
    \midrule
    \multirow{3}{*}{768x768x768} & \multirow{3}{*}{Float} &6(100\%)	&\textbf{0.44}&	\textbf{2263.35}&\textbf{2050.53}&5.34&33.02&62.10\\
    ~ & ~ & 3(50\%)&0.82&1216.01&1101.67&5.74&17.54&62.81\\
    ~ & ~ & 1(17\%)&1.84&542.62&491.60&\textbf{7.68}&7.42&66.25\\
    \midrule
    \multirow{3}{*}{1536x1536x1536} & \multirow{3}{*}{Float} &6(100\%)	&\textbf{2.41}&	\textbf{415.11}&\textbf{3008.63}&7.83&39.39&76.38\\
    ~ & ~ & 3(50\%)&4.45&224.82&1629.45&8.49&21.08&77.30\\
    ~ & ~ & 1(17\%)&12.99&76.99&558.02&\textbf{8.72}&7.88&70.81\\
    \midrule
    \multirow{3}{*}{3072x3072x3072} & \multirow{3}{*}{Float} &6(100\%)	&\textbf{17.17}&\textbf{58.25}&\textbf{3377.66}&8.80&41.8&80.67\\
    ~ & ~ & 3(50\%)&34.12&29.31&1699.19&8.85&21.55&78.85\\
    ~ & ~ & 1(17\%)&101.82&9.82&569.44&\textbf{8.90}&7.96&71.54\\
    \midrule
    \multirow{3}{*}{6144x6144x6144} & \multirow{3}{*}{Float} &6(100\%)	&\textbf{135.59}&\textbf{7.38}&\textbf{3421.02}&8.90&42.13&\textbf{81.20}\\
    ~ & ~ & 3(50\%)&270.85&3.69&1712.61&\textbf{8.92}&21.65&79.10\\
    ~ & ~ & 1(17\%)&812.13&1.23&571.16&\textbf{8.92}&7.97&71.66\\
    \bottomrule
  \end{tabular}
  }
\end{table*}

Table 6 shows the performance of the MM accelerator under different task scales and different numbers of PUs. The peak performance of the MM accelerator under the Float data type is 3421.02 GOPS, the maximum power consumption is 42.13 W, and the energy efficiency at the peak performance is 81.20 GOPS/W. If the low bit types such as Int8 or Int16 are used, higher energy efficiency will be obtained, which has huge advantages over the GPU. Because the selected task scale is large, it can meet the simultaneous operation of all PUs, so similar GOPS can be obtained under different task scales, and obvious performance gap can be seen when adjusting the number of PUs. In addition, The MM accelerator can maintain a high AIE single core performance. Although the AIE single core performance decreases with the increase of the number of PU in smaller MMs, when the task scale continues to expand, The AIE single core performance gap caused by the PU quantity adjustment has disappeared, and the highest AIE single core efficiency obtained is 8.92 GOPS.

\begin{table*}[t]
  \caption{Performance of the Filter2D accelerator at different task scales and PU quantities.}
  \label{tab:commands}
  \scalebox{0.9}{ %set scale
  \begin{tabular}{cccc cccc cc}
    \toprule
    Problem Size&\makecell{Data\\Type}& \makecell{PU\\Quantity}&\makecell{Time\\(ms)}&Tasks/sec&GOPS&GOPS/AIE&\makecell{Power\\(W)}&GOPS/W\\
    \midrule
    \multirow{3}{*}{128x128,5x5} & \multirow{3}{*}{Int32} &44(100\%)&	\textbf{0.15}&	\textbf{6468.72}&	\textbf{5.30}&	0.015&15.29&0.35\\
    ~ & ~  & 20(45\%)&	0.16&	6354.41&	5.21&	0.03&14.18&0.37\\
    ~ & ~  & 4(9\%)&	0.16&	6176.00&	5.06&	\textbf{0.158}&6.45&0.78\\
    \midrule
    \multirow{3}{*}{3480x2160(4K),5x5} & \multirow{3}{*}{Int32} &44(100\%)&	\textbf{0.43}&	\textbf{2315.94}&\textbf{870.42}&2.472&28.29&30.77\\
    ~ & ~  & 20(45\%)&	0.91&	1100.91&	413.76	&2.586&22.59&18.31\\
    ~ & ~  & 4(9\%)&	3.91&	255.80&	96.14&	\textbf{3.004}&8.31&11.57\\
    \midrule
    \multirow{3}{*}{7680x4320(8K),5x5} & \multirow{3}{*}{Int32} &44(100\%)&	\textbf{1.67}&	\textbf{595.92}&	\textbf{988.56}	&2.808&30.20&\textbf{32.73}\\
    ~ & ~  & 20(45\%)&	3.51&	284.58&	472.10&	2.951&23.79&19.84\\
    ~ & ~  & 4(9\%)&	17.04&	58.69&	97.37&	\textbf{3.042}&8.33&11.69\\
    \midrule
    \multirow{3}{*}{15360x8640(16K),5x5} & \multirow{3}{*}{Int32} &44(100\%)&\textbf{	6.32}&	\textbf{158.30}&	\textbf{1050.43}&2.984&35.62&29.49\\
    ~ & ~  & 20(45\%)&	13.71&	72.94&	484.02&	3.025&24.04&20.13\\
    ~ & ~  & 4(9\%)&	67.73&	41.76&	97.97&	\textbf{3.061}&8.35&11.73\\
     
    \bottomrule
  \end{tabular}
  }
\end{table*}

Table 7 shows the performance of the Filter2D accelerator under different task scales and different PU quantities. The peak performance reaches 1050.43GOPS, and the maximum energy efficiency is 32.73GOPS/W. It can be seen that the Filter2D accelerator can achieve high performance under the condition of adaptive resolution. When calculating the smaller resolution, because the split task size is 32x32 image blocks, it cannot use all the PUs for calculation, and even if more PUs are allocated, it cannot be further accelerated. Under the 128x128 resolution, there is only a maximum of 5.30 GOPS, but the performance of the Tasks/sec parameter is outstanding. In other higher resolution tests, all the PUs can participate in the calculation, so it can see obvious performance differences when reducing the number of PUs. Limited by PL performance, serving more PUs will reduce the AIE single core efficiency. However, with the increase of task size, the AIE single core efficiency obtained under different PU quantities tends to be the same.

\begin{table*}[t]
  \caption{Performance of the FFT accelerator at different task scales and PU usage quantities.}
  \label{tab:commands}
   \scalebox{0.9}{ %set scale
  \begin{tabular}{cccc ccc}
    \toprule
    Sample Size&	Data Type&	PU Quantity&	Run Time (us)&	Tasks/sec& Power(W)&Tasks/sec/W\\
    \midrule
    \multirow{3}{*}{8192} & \multirow{3}{*}{CInt16} &8(100\%)&	\textbf{4.00}&	\textbf{250000.00}&12.03&20781.38\\
    ~ & ~ & 4(50\%)&	8.12&	123152.71&10.38&11864.43\\
    ~ & ~ & 2(25\%)&	N/A&	N/A&N/A&N/A\\
    \midrule
    \multirow{3}{*}{4096} & \multirow{3}{*}{CInt16} &8(100\%)&	\textbf{1.90}&	\textbf{526315.79}&12.26&42929.51\\
    ~ & ~ & 4(50\%)&	4.00&	250000.00&10.44&23946.36\\
    ~ & ~ & 2(25\%)&	7.49&	133511.35&9.56&13965.62\\
    \midrule
    \multirow{3}{*}{2048} & \multirow{3}{*}{CInt16} &8(100\%)&	\textbf{0.89}&	\textbf{1123595.51}&12.47&90103.89\\
    ~ & ~ & 4(50\%)&	1.73&	578034.68&9.87&58564.81\\
    ~ & ~ & 2(25\%)&	3.62&	276243.10&9.62&28715.50\\
    \midrule
    \multirow{3}{*}{1024} & \multirow{3}{*}{CInt16} &8(100\%)&	\textbf{0.43}&	\textbf{2325581.40}&12.58&184863.39\\
    ~ & ~ & 4(50\%)&	0.85&	1176470.59&9.90&118835.41\\
    ~ & ~ & 2(25\%)&	1.70&	588235.29&9.74&60393.77\\
     
    \bottomrule
  \end{tabular}
  }
\end{table*}

In the FFT application, we tested the performance of FFT accelerator under different sample points and different PU quantities. Because FFT has many communications, it is difficult to accurately reflect the overall performance by GOPS indicators alone, so we used Tasks/sec as the performance evaluation indicator and Tasks/sec/W(TPS/W) as the energy efficiency indicator. Table 8 shows the performance of the FFT accelerator under different task sizes and different processing units. Because the size of the data and intermediate results involved in the FFT calculation of 8192 samples exceeds the memory size of the AIE calculation core of two PUs, this sample point is only applicable to the configuration of four or eight PUs.

Since FFT operator only uses 80 AIE cores, in actual application scenarios, multiple groups of such FFT accelerators can be deployed in the board for parallel operation, which can achieve higher performance and improve hardware utilization.

\begin{table*}[t]
  \caption{Test results of performance testing of AIE computing based on MM.}
  \label{tab:commands}
  \scalebox{0.9}{ %set scale
  \begin{tabular}{cccc cccc}
    \toprule
     ID&	Data Type&	AIE frequency&	Tasks/sec&	GOPS&	GOPS/AIE& Power(W) & GOPS/W\\
    \midrule
    1&	Float&	1.33GHZ&	$9.37\times10^7$&	6141.51&15.35&65.17&94.24\\
    2&	Float&	1.33GHZ&	$9.53\times10^7$&	\textbf{6243.29}&\textbf{15.61}&66.27&94.21\\
    3&	Float&	1.33GHZ&	$9.39\times10^7$&	6159.89&15.40&65.38&94.21\\
    Average&	N/A&N/A&	$9.43\times10^7$&	6181.56&15.45&65.61&94.22\\
    \bottomrule
  \end{tabular}
  }
\end{table*}

As an AIE performance test program, MM-T can minimize the performance loss caused by communication and other factors to ensure the accuracy of the test. Mm-t uses 32 × 32 × 32 MM of the Float data type as the base task, and the compute engine uses all 400 AIE cores. Table 9 shows the results of the MMT test. Using the method of averaging three tests as the final test result, the results show that when AIE is at the highest clock frequency, the average $9.43\times10^7$ MM operations are completed per second, the average GOPS is 6181.56, and the average energy efficiency is 94.22GOPS/W.

\subsection{Performance comparison}

\begin{table*}[h]
  \caption{Performance and energy efficiency comparison between EA4RCA framework and current SOTA design.}
  \label{tab:commands}
  \scalebox{0.8}{ %set scale
  \begin{tabular}{cccc cccc c}
    \toprule
    Apps&
    Design&
    \makecell[c]{Problem\\Size}&
    \makecell[c]{Data\\Type}&
    \makecell[c]{Task/sec\\(TPS)}&
    GOPS&
    \makecell[c]{Energy\\Efficiency}&
    \makecell[c]{Speed\\ Up Ratio}&
     \makecell[c]{Energy\\Efficiency\\ Up Ratio}\\
    \midrule
    
    \multirow{2}{*}{MM} &CHARM\cite{acap_impl3}&N/A&Float&N/A&3270.00&62.40 GOPS/W&1.00x&1.00x\\
    ~& EA4RCA&6144&Float&7.38&\textbf{3421.02}& \textbf{81.20 GOPS/W} &\textbf{1.05x}&\textbf{1.30x}\\
    \midrule
    
    \multirow{7}{*}{Filter2D}&CCC2023\cite{CCC2023}&\makecell[c]{4K\\(3x3)}&Int32&289.32&39.22&5.04 GOPS/W&1.00x&1.00x\\
    ~ & CCC2023\cite{CCC2023}&\makecell[c]{8K\\(3x3)}&Int32&98.78&59.72&7.68 GOPS/W&1.00x&1.00x\\
    ~ & EA4RCA&	\makecell[c]{4K\\(5x5)}&Int32&2315.94&\textbf{870.42}&\textbf{30.77 GOPS/W}&\textbf{22.19x}&\textbf{6.11x}\\
    ~ & EA4RCA&	\makecell[c]{8K\\(5x5)}&Int32&595.92&\textbf{988.56}&\textbf{32.73 GOPS/W}&\textbf{16.55x}&\textbf{4.26x}\\
    \midrule
    
    \multirow{7}{*}{FFT} &Vitis\cite{vitis_lib}&	1024&	CInt16&	713826.80&	N/A&N/A&1.00x&N/A\\
    ~& CCC2023\cite{CCC2023}&1024&CInt16&142857.14&N/A&26396.37 TPS/W&0.20x&1.00x\\
    ~& CCC2023\cite{CCC2023}&4096&CInt16&135685.21&N/A&22796.57 TPS/W&1.00x&1.00x\\
    ~& CCC2023\cite{CCC2023}&8192&CInt16&106382.97&N/A&16396.88 TPS/W&1.00x&1.00x\\
    ~& EA4RCA&1024&CInt16&\textbf{2325581.40}&N/A&\textbf{184863.4 TPS/W}&\textbf{3.26x}&\textbf{7.00x}\\
    ~& EA4RCA&4096&CInt16&\textbf{526315.79}&N/A&\textbf{42929.51 TPS/W}&\textbf{3.88x}&\textbf{1.88x}\\
    ~& EA4RCA&8192&CInt16&\textbf{250000.00}&N/A&\textbf{20781.38 TPS/W}&\textbf{2.35x}&\textbf{1.27x}\\
    \midrule
    
     \multirow{2}{*}{MM-T}&CHARM\cite{acap_impl3}&N/A&Float&N/A&3270.00&62.40 GOPS/W&1.00x&1.00x\\
    ~& EA4RCA&32&Float&$9.43\times10^7$&\textbf{6181.56}&\textbf{94.22 GOPS/W}&\textbf{1.89x}&\textbf{1.51x}\\

    \bottomrule
  \end{tabular}
  }%scale
\end{table*}

In order to further demonstrate the advantages of the accelerator designed based on the EA4RCA framework in all aspects, this paper compares the performance, energy efficiency and adaptive task scale with the highest performance accelerator (SOTA) designed based on the VCK5000 platform. The comparison results are shown in Table 10.

The peak performance of the MM accelerator designed based on the EA4RCA framework is 3421.02 GOPS, and the energy efficiency is 81.20 GOPS/W, which is 1.05x and 1.30x of the current SORM scheme CHARM\cite{acap_impl3}. In addition, our MM accelerator is more scalable and flexible.

The Filter2D accelerator was compared with the scheme of the champion team of the Custom Computing Algorithm Challenge (CCC2023)\cite{CCC2023}. The performance at 4K and 8K resolution was improved to 22.19x and 16.55x, and the energy efficiency was improved to 6.11x and 4.26x. At the same time, we also support task scale adaptation and dynamic PU quantity adjustment, which is impossible for the current SOTA scheme.

The FFT accelerator compares the current SOTA scheme with three sample sizes of 1024, 4096 and 8192. The performance improvement is 3.26x, 3.88x and 2.35x respectively, and the energy efficiency improvement is 7.00x, 1.88x and 1.27x respectively. Unlike other accelerators, the energy efficiency of FFT accelerator is evaluated by Tasks/sec/W (TPS/W), because the throughput is difficult to comprehensively display the performance of FFT accelerator. In addition, our FFT accelerator also supports such features as task scale adaptation.

For MM-T, due to the particularity of its design purpose, there is no performance impact of peripheral hardware, so its peak performance reaches 6181.56 GOPS, and the energy efficiency under the peak performance is 94.22 GOPS/W, 1.81x the performance of MM experiment in this paper, 1.89x the CHARM experiment, 1.16x the energy efficiency of MM experiment in this paper, 1.51x the CHARM experiment, reaching the current highest level.

\section{Conclusion}
This paper addresses the challenges associated with underutilization of powerful and flexible hardware resources and low development efficiency during accelerator deployment in the ACAP architecture. To address these challenges, we propose the EA4RCA framework, which follows a design pattern based on focused computation and unified communication, tailored to the application characteristics. The framework adopts a "top-down" approach, beginning with the design of the AIE and subsequently constructing the deployment process for peripheral services based on AIE requirements. This approach allows for adaptation and high-performance capabilities for various applications.

In order to achieve efficient computing and data supply and reception, The EA4RCA framework adopts a highly configurable and extensible AIE processing unit design, as well as a high-performance data engine for AIE, which supports task scale adaptation and parallel data flow services. In addition, it is equipped with AIE Graph code generator to improve the efficiency of AIE development. Experiments show that compared with other accelerator implementations, the scheme in this paper has certain advantages in computing resource utilization, task scale adaptation, and development efficiency; Based on Xilinx VCK5000 Versal development platform, the EA4RCA framework is used to realize MM The efficiency of Filter2D, FFT and MM-T has obvious advantages over other implementation schemes. The highest throughput reaches 6181.56 GOPS, and the highest energy efficiency is 94.22 GOPS/W, which is the highest level at present. These experiments prove that this framework can achieve high parallelism and improve development efficiency when deploying accelerators.

The focus of this paper is on the efficient deployment of accelerators in the ACAP architecture, specifically highlighting the unique high-performance computing characteristics of the AIE. The key design aspect involves leveraging communication patterns for parallelization. Our exploratory work reveals that RCA-class applications can effectively separate the computation and communication processes, aligning with the suitability of the AIE architecture for parallel optimization design. Consequently, optimizing the data flow design within the system resource constraints in the AIE architecture is adequate for fully leveraging its maximum performance.

Based on the above work, on the one hand, CA algorithms can be further subdivided to build a more complete optimization implementation framework based on AIE accelerator in the future; On the other hand, it can expand the fully automated deployment framework with AIE as the core under different engineering demand indicators, and promote such processors to exert their energy in their advantageous areas.

\section{Acknowledgments}
We thank the Versal VCK 5000 board donated by the Xilinx University Program (XUP). We thank UCLA's AMD/Xilinx heterogeneous accelerated computing cluster.

\bibliographystyle{ACM-Reference-Format}
\bibliography{sample-base}

\appendix

\end{document}